\documentclass[12pt]{iopart}
\usepackage[dvips]{graphicx}
\usepackage{color}
 \expandafter\let\csname equation*\endcsname\relax
 \expandafter\let\csname endequation*\endcsname\relax 
\usepackage{amsfonts}
\usepackage{amsmath}
\usepackage{textcomp}

\usepackage{verbatim}
\usepackage{ulem}
\usepackage{upgreek}
\usepackage{epstopdf}
\usepackage{multirow}
\usepackage{eqnarray}
\usepackage{setspace}
\usepackage{ulem}
\usepackage{iopams}
\usepackage{sidecap}
\begin{document}

\title[]{Measurements of the exchange stiffness of YIG films by microwave resonance techniques}
\author{S~Klingler$^1$, A~V~Chumak$^1$, T~Mewes$^2$, B~Khodadadi$^2$, C~Mewes$^2$, C~Dubs$^3$, O~Surzhenko$^3$, B~Hillebrands$^1$, A~Conca$^1$}
\address{$^1$Fachbereich Physik and Landesforschungszentrum OPTIMAS, Technische Universit\"at Kaiserslautern, 67663 Kaiserslautern, Germany}
\address{$^2$Department of Physics and Astronomy, MINT Center, University of Alabama, Tuscaloosa, Alabama 35487, USA}
\address{$^3$INNOVENT e.V. Technologieentwicklung, 07745 Jena, Germany}
\ead{klingler@physik.uni-kl.de}
\date{\today}

\begin{abstract}
Measurements of the exchange stiffness $D$ and the exchange constant $A$ of Yttrium Iron Garnet (YIG) films are presented. The YIG films with thicknesses from 0.9\,$\mu$m to 2.6\,$\mu$m were investigated with a microwave setup in a wide frequency range from 5 to 40\,GHz. The measurements were performed when the external static magnetic field was applied in-plane and out-of-plane. The method of Schreiber and Frait \cite{frait}, based on the analysis of the perpendicular standing spin wave (PSSW) mode frequency dependence on the applied out-of-plane magnetic field, was used to obtain the exchange stiffness $D$. This method was modified to avoid the influence of internal magnetic fields during the determination of the exchange stiffness. Furthermore, the method was adapted for in-plane measurements as well. The results obtained using all methods are compared and values of \textit{D} between $(5.18\pm0.01) \cdot 10^{-17}$\,T$\cdot$m$^2$ and $(5.34\pm0.02) \cdot 10^{-17}$\,T$\cdot$m$^2$ were obtained for different thicknesses. From this the exchange constant was calculated to be $A=(3.65 \pm 0.38)$\,pJ/m.
\end{abstract}

\maketitle

\section{Introduction}

In order to employ the degree of freedom of the spin in future information technology, materials with low Gilbert damping and long spin-wave propagation distances are needed for data transport. Yttrium Iron Garnet (YIG) is a material which fulfills the aforesaid requirements. New technologies employing YIG are being developed and new physical phenomena were investigated. Logic operations with spin waves in YIG waveguides \cite{Khitun, Schneider, Klingler}, data-buffering elements \cite{Chumak2012} and magnon transistors \cite{Chumak} are only a few examples for the latest technology progress. Especially, YIG films of nanometer thickness \cite{Jungfleisch,pirro,  Fert, Wu1, Wu2, Hahn} are of large importance since they allow for the realization of nano- and microstructures \cite{Chumak, pirro, Hahn2014, Florin} and an enhancement of spin-transfer-torque related effects \cite{Hahn, Kajiwara}. In this context the material parameters of YIG are of crucial importance for its application potential. 

In a magnetic system, the exchange interaction contributes strongly to the energy of the system. From a classical point of view, this interaction is responsible for the parallel alignment of adjacent spins, thus, it strongly influences the spin-wave characteristics. The strength of the exchange interaction is given by the exchange stiffness $D$, but the existing approaches for its measurement are often influenced by internal magnetic fields depending consequently on crystal anisotropies and the saturation magnetization.
Thus, methods are required for the exact determination of the exchange stiffness without the uncertainties added by the aforementioned parameters. Here, such a method is presented and compared to the results obtained by commonly used data evaluation methods. Firstly, the classical approach of Schreiber and Frait \cite{frait} is used for the determination of the exchange stiffness when the external static magnetic field is applied out-of-plane. Secondly, the method is modified to avoid any influence of the anisotropy fields and the saturation magnetization in order to achieve highly-precise values for $D$. Thirdly, the method of Schreiber and Frait is adapted and used for in-plane measurements. All values of $D$ obtained by using the different methods are compared and, then, values of the exchange constant $A$ are calculated for our YIG samples. 

\section{Theory}
The precessional motion of the magnetization in an effective magnetic field is described by the Landau-Lifshitz and Gilbert equation \cite{Landau-Lifshitz}. The effective magnetic field depends on various parameters, such as the applied static  and time dependent  magnetic fields ($\mu_0 H_{0}$ and $\mu_0 h(t)$, respectively), anisotropy fields (the cubic anisotropy field  $\mu_0 H_\mathrm{c}$, the uniaxial out-of-plane $\mu_0 H_\mathrm{u\perp}$ and in-plane $\mu_0 H_\mathrm{u\parallel}$ anisotropy fields), as well as the exchange field $\mu_0 H_\mathrm{ex}=Dk^2$ which describes the exchange interaction in the investigated material. Here, $D=\frac{2A}{M_\mathrm{S}}$ is the exchange stiffness, $A$ the exchange constant and $M_\mathrm{S}$ the saturation magnetization. The wavevector $k$ is the wavevector of perpendicular standing spin-waves which is quantized over the sample thickness. Under the assumption of perfect pinning of the spins at the sample surface $k$ is defined by $k=n\pi/d$ \cite{frait}, where $n$ is the mode number. The case $n=0$ corresponds to the classical case of ferromagnetic resonance. 

The resonant precession frequency for cubic crystals is presented in reference \cite{Bobkov}. For the case when the static magnetic field is applied in-plane one obtains
		\begin{equation}
		\label{klingler:eq:kittel_ip}
		\left( \frac{\omega_\parallel}{|\gamma|}\right)^2 =  \mu_0^2\left(H_\mathrm{0}+H_\mathrm{ex}\right)  \left(H_\mathrm{0}+H_\mathrm{ex}+ M_\mathrm{S}- H_\mathrm{u\perp}-H_\mathrm{c} \right).
		\end{equation}
This equation is valid if the magnetization of the sample points along the $\langle110\rangle$-axis of the crystal.
If the static magnetic field is applied out-of-plane the frequency is given by
		\begin{equation}
		\label{klingler:eq:kittel_oop}
		\frac{\omega_\perp}{|\gamma|}=\mu_0 \left(H_\mathrm{0} + H_\mathrm{ex}  -  M_\mathrm{S}+  H_\mathrm{u\perp}-\frac{4}{3} H_\mathrm{c}+H_\mathrm{u\parallel}\right). 
		\end{equation}
Here, $\omega_\perp$ and $\omega_\parallel$ are the applied microwave frequencies, $\gamma$ is the gyromagnetic ratio and $\mu_0 H_0$ is the applied static magnetic field.

\section{Samples and Experimental Setup}

\begin{table}
 \caption{\label{dubs:samples} Parameters of the studied YIG samples. The average growth rate $\nu$ was calculated from the thickness $d$ and the deposition time which was 5\,min for all YIG films.}
  \renewcommand{\arraystretch}{1.1}% Wider 
 \begin{indented}

 \item[]\begin{tabular}{@{}cccc}
\mr
\multirow{2}{*}{Sample}	& Thickness		&	Growth rate 	&	Lattice misfit \\
		& $d$\,($\mu$m)		&	$\nu$\,($\mu$m/min)&	$\Delta a^\perp/a_\mathrm{GGG}\,(10^{-4})$ \\
 \mr
E1		& $2.59\pm0.01$		&	0.52		&	$+5.33\pm0.07$ \\
E2		& $1.59\pm0.02$		&	0.32		&	$+7.68\pm0.02$ \\
E3		& $0.903\pm0.003$	&	0.18		&	$+8.72\pm0.03$ \\
 \mr
 \end{tabular}
 \end{indented}
 \end{table}

The YIG films were grown by liquid phase epitaxy (LPE) on (111)-oriented Gadolinium Gallium Garnet (GGG) substrates. Due to the difference in the lattice parameters of Czochralski-grown GGG (a$_\mathrm{GGG}=12.383\,$\AA) and pure YIG (a$_\mathrm{YIG}=12.376\,$\AA) \cite{Wang} the films exhibit a room temperature lattice misfit $\Delta a^{\perp}=a_\mathrm{GGG}-a_\mathrm{YIG}$ which results in strained epitaxial films. This strain is one of the main factors defining the uniaxial anisotropy fields $\mu_0 H_\mathrm{u\perp}$ and $\mu_0 H_\mathrm{u\parallel}$. In the case of LPE growth of garnet films incorporation of lead ions from the PbO solvent plays an important role in adjusting the film misfit \cite{Hergt}. Therefore, the misfit essentially depends on growth parameters (growth temperature, growth rate, etc.). For this reason, the growth rate $\nu$ was varied to obtain (Y$_{1-x}$Pb$_x$)$_3$Fe$_5$O$_{12}$ films ($0.005\leq x\leq 0.015$ \cite{DubsExplain}) with reduced lattice misfits. In Tab.~\ref{dubs:samples} important material parameters of the samples are shown. It can be seen that the lattice misfit increases with decreasing growth rate. The film thickness $d$ was measured by a prism coupler technique, and the YIG/GGG lattice misfit values were determined by X-ray diffraction. Then, the samples were cut in sizes of $3 \times 3$\,mm$^2$ for microwave studies.

For measuring the exchange stiffness, a waveguide microwave resonance setup was used. An electromagnet is used to apply external fields up to $\mu_0 H_{dc}<1650\,$mT$\pm 0.1$\,mT, where a low-amplitude ($\mu_0 H_{ac}=0.1\,$mT) rf-frequency ($f\ll1\,$MHz) modulation field is used by a lock-in amplifier as reference signal. The scan of a Lorentzian absorption peak with the modulation field results in an output voltage which has the form of the derivative of the original signal. A microwave field with a power of $10\,$dBm is applied in a wide frequency range from 5\,GHz to 40\,GHz with a rotatable coplanar waveguide (CPW) so that the angle between the field and the sample surface can be varied from 0$^\circ$ to 360$^\circ$. For the in-plane measurements the external magnetic field is applied along the edges of the sample which is positioned in the middle of the CPW. In all measurements the frequency is fixed and the field is swept.

 \section{Determination of exchange stiffness}
 \subsection{Method of Schreiber and Frait}

\begin{figure}[t!]
 	  \begin{center}
     \scalebox{1}{\includegraphics[width=1\linewidth, clip]{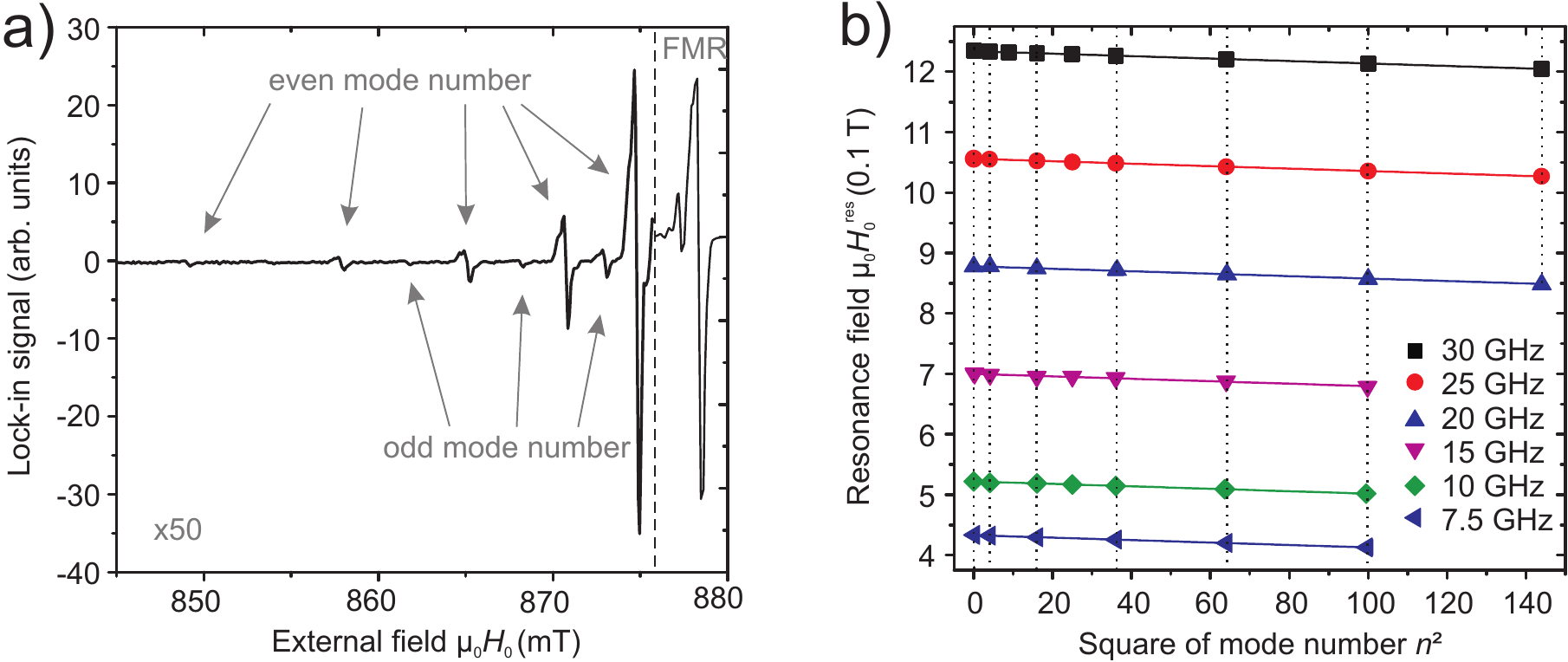}}
     \end{center}
 	  \caption{\label{klingler:fig:spectrum} (a) Example spectrum for sample E2 in the out-of-plane configuration at 20\,GHz. The spectrum on the left hand side of the dashed lines is magnified by a  factor of 50. (b) Line positions found in the spectrum of sample E2 in out-of-plane configuration. It is obvious that the line positions follow a linear function in dependence of $n^2$. Since $D$ is a shared fitting parameter, it is the same for every frequency. The dashed lines show the positions of the resonances with even mode numbers.}
 \end{figure}

A typical dependence of the lock-in signal on the applied static field from the out-of-plane measurements is shown in Fig. \ref{klingler:fig:spectrum}(a). The ferromagnetic resonance ($n=0$) can be found at the highest field values, whereas the thickness modes are located at lower field values. In any case mainly resonances with an even mode number are observed. This effect can be understood with the assumption of ``perfect pinning'', since in this case, only the even modes absorb energy from the homogeneous antenna field \cite{Kittel1958}. 
The experimental observation of odd PSSWs can be caused by small microwave inhomogeneities across the film thickness.

In the classical approach of Schreiber and Frait the exchange stiffness is determined in the out-of-plane configuration using Eq.~(\ref{klingler:eq:kittel_oop}). Here, the anisotropy fields and the saturation magnetization are absorbed in the effective magnetization $M_\mathrm{eff,\perp}$. Then, the resonance field for a certain frequency is defined by:
	\begin{gather}
	\label{eq:exchange}
	 \mu_0 H_\mathrm{0}^\mathrm{res}(n)=\mu_0 M_\mathrm{eff,\perp}+\frac{\omega_\perp}{|\gamma|}-D \frac{\pi^2}{d^2}n^2\text{, where}\\
	 \label{eq:exchange2}
	 M_\mathrm{eff, \perp}=M_\mathrm{S} -H_\mathrm{u\perp}+\frac{4}{3}H_\mathrm{c}-H_\mathrm{u\parallel}	\end{gather}
	
In a plot where the resonance field is drawn over the square of the mode number $n^2$, the exchange stiffness can be extracted with the slope of a linear function, where the $y$-intercept delivers information about the effective magnetization. The presence of resonances with odd mode numbers introduces some ambiguity regarding the identification of the modes. However, the mode intensity together with the $n^2$-dependence of the resonance field shift enables a consistent identification, as can be seen in Fig~\ref{klingler:fig:spectrum}(b). Here, the resonance fields of sample E2 are shown for different frequencies. The slopes of the linear functions are the same for all measurements and the $y$-intercepts are different due to the use of different excitation frequencies. In the performed measurements no deviations from the linear functions were detected which would occur for small $n$ due influence of the surface anisotropy. Thus, the assumption of perfect pinning is justified. From the slope the exchange stiffness values from all samples are extracted, which are presented in the left column of Tab.~\ref{klingler:tab:results}. The average value for samples E1-E3 is $D=(5.32 \pm 0.07)\times10^{-17}\,\text{T}\cdot \text{m}^2$. With the shown method, the slope and the effective magnetization are optimized together during the fitting process, i.e. the residuum is minimized. The optimization of both parameters at the same time leads to a mutual influence of the parameters. This effect is clearly visible in the size of the error bars, if compared to the modified method which is presented in the next section.

 \begin{table}
 \caption{\label{klingler:tab:results} Results for the YIG samples with different thicknesses. The shown errors are the statistical fitting errors. The values in the column with the out-of-plane$^\dagger$ measurement are obtained using the original method of Schreiber and Frait. The values in the column with the out-of-plane$^*$ measurements are obtained based on the difference between the resonance field of higher modes and the ferromagnetic resonance field.}
 
 \begin{indented}
 \renewcommand{\arraystretch}{1.1}% Wider 
 \item[]\begin{tabular}{@{}cccc}
 \mr
  & \multicolumn{3}{c}{\textit{D}\,$(10^{-17}$\,T$\cdot$m$^2)$ or $10^{-9}$\,(erg/G$\cdot$cm)} \\
   	{Sample }		 	 &\textit{out-of-plane$^\dagger$} & \textit{out-of-plane$^*$} &\textit{in-plane}\\
   & \footnotesize{Schreiber and Frait} & & \\
 \mr
  E1& $5.33 \pm 0.09$	& $5.18 \pm 0.01$ & $5.29 \pm 0.04$	\\
  E2 & $5.32 \pm 0.09$	& $5.34 \pm 0.02$ & $5.30 \pm 0.02$ \\
  E3 & $5.29 \pm 0.05$ 	& $5.31 \pm 0.02$ & $5.40 \pm 0.02$	\\
 \mr
 \end{tabular}
 \end{indented}
 \end{table}

 \begin{figure}[t!]
 	  \begin{center}
     \scalebox{1}{\includegraphics[width=0.5\linewidth, clip]{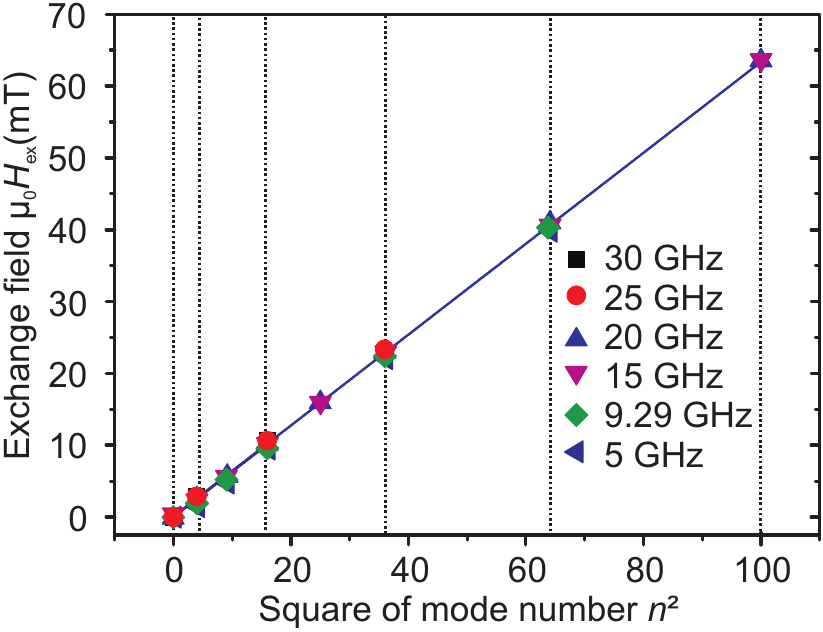}}
     \end{center}
 	  \caption{\label{klingler:fig:highermodes-fmr} The plotted exchange fields of sample E3 are obtained building the difference between the resonance fields of the higher modes ($n \neq 0$) and the ferromagnetic resonance ($n=0$) field. The exchange fields only depend on the square of the mode number and not on the frequency.}
 \end{figure}

\subsection{Modified method of Schreiber and Frait}

As shown before, the method of Schreiber and Frait requires several parameters to be taken into account in order to obtain the exchange stiffness. Here, a method which is completely independent on assumptions for the anisotropy fields and the saturation magnetization is presented. For this the ferromagnetic resonance field $\mu_0 H_0^{\mathrm{res}}(0)$ is subtracted from the resonance fields of the higher modes $\mu_0 H_0^{\mathrm{res}}(n\neq0)$ in order to determine the exchange field $\mu_0 H_\mathrm{ex}$ of the thickness modes. Since the resonance field of the ferromagnetic resonance contains all information about the anisotropy fields and $M_\mathrm{S}$, as can be seen in Eq.~(\ref{eq:exchange2}), the exchange field only depends on $D$:
 	\begin{equation}
 	\label{eq:exchange_contribution}
 	\mu_0 H_\mathrm{ex}=\mu_0 H_\mathrm{0}^\mathrm{res}(n)-\mu_0 H_\mathrm{0}^\mathrm{res}(0)=D \frac{\pi^2}{d^2}n^2
 	\end{equation} 	
In Fig.~\ref{klingler:fig:highermodes-fmr} the exchange fields are shown as a function of $n^2$. One can see that the measured exchange fields for different frequencies collapse in a point for each mode number. This indicates that the exchange fields are independent on any external parameter. Furthermore, all collapsed data points lie on a linear function with $H_\mathrm{ex}(0)=0$. This data can now be analyzed using a simple linear fit with no offset, i.e. only one fitting parameter is used. Thus, any mutual influence of parameters is avoided which is the reason for a significantly reduced statistical error. The results are shown in the middle column of Tab.~\ref{klingler:tab:results}. All values are in the same range as obtained with the former method. However, it is visible that the exchange stiffness of sample E1 is significantly smaller that the others. This difference can be understood by a larger saturation magnetization for sample E1 than for samples E2 and E3 as shown below. In comparison to the method of Schreiber and Frait, the error is decreased by a factor of up to 9 due to the avoided influence of the effective magnetization during the data evaluation. This allows for the identification of the exchange stiffness with a high accuracy.

\subsection{Method for in-plane measurements}

 \begin{figure}[t!]
 	  \begin{center}
     \scalebox{1}{\includegraphics[width=1\linewidth, clip]{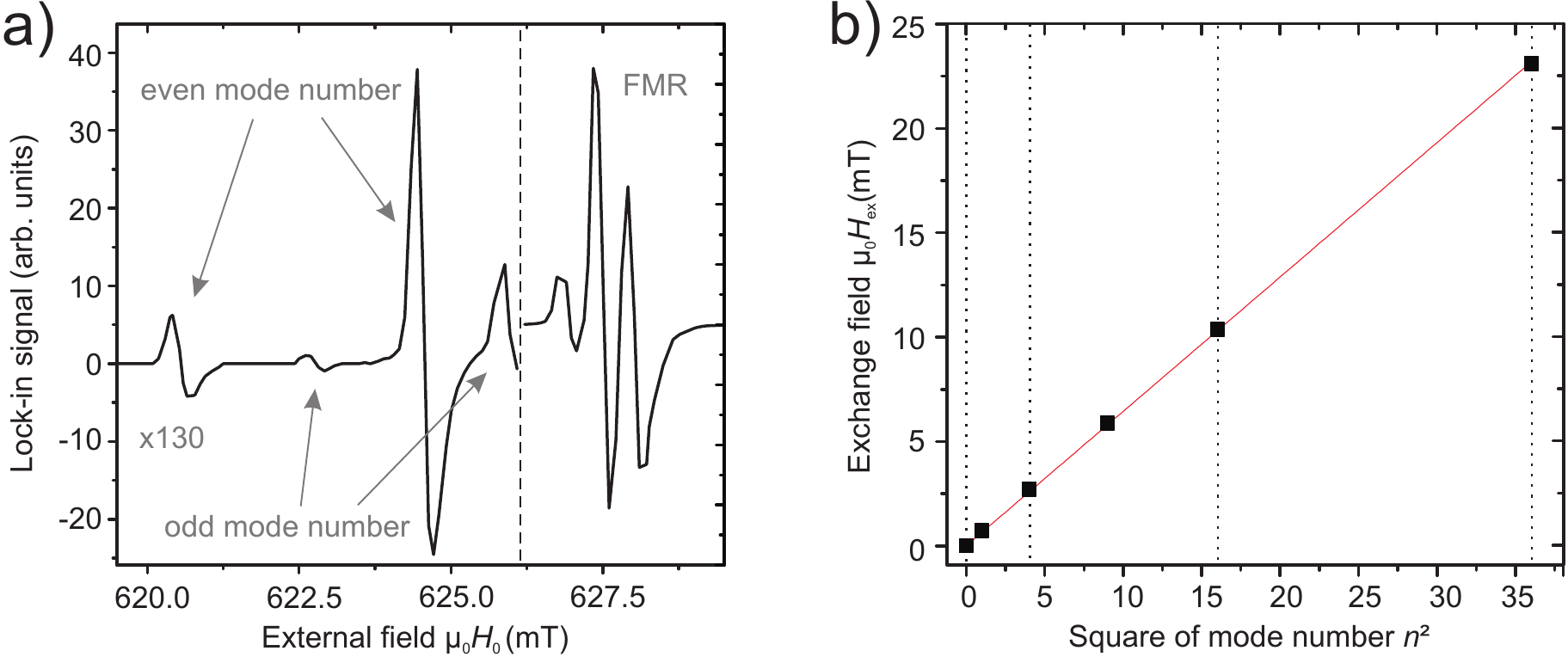}}
     \end{center}
 	  \caption{\label{klingler:fig:spectrum2} (a) Example spectrum for the YIG sample E2 in in-plane configuration at 20\,GHz. The first two resonances overlap in such a way that a multiple resonance fit had to be used to extract the linewidth and position of both resonances. The spectrum on the left hand side of the dashed lines is magnified by a  factor of 130. (b) The exchange fields of the sample E3 follow a linear function of the square of the mode number. The slope of the function is proportional to the exchange constant. Each point in the graph stands for the exchange field of one resonance. Even modes are marked with dashed lines.}
 \end{figure}

Classically the method of Schreiber and Frait is used for the determination of the exchange stiffness in out-of-plane configuration. Here, the method is adapted for the use in in-plane configuration. A sample spectrum of the in-plane measurements is shown in Fig.~\ref{klingler:fig:spectrum2}(a). It is slightly modified in comparison to the out-of-plane spectrum. The resonances are shifted to smaller field values due to decreased demagnetizing effects. In the in-plane case, the former methods cannot be used for data evaluation since $\omega_\parallel$ is not linearly dependent on the static field in Eq. (\ref{klingler:eq:kittel_ip}). However, the former procedure can be applied to the pure exchange field of the PSSWs. For this we propose the following steps.

Firstly, Eq.~(\ref{klingler:eq:kittel_ip}) must be rewritten in a way which is convenient for the fitting process:
 	\begin{equation}
 	\omega_\mathrm{\parallel}=|\gamma|\mu_0\sqrt{\left( H_0+ H_\mathrm{ex}\right)\left(H_0+H_\mathrm{ex}+ M_\mathrm{eff, \parallel}\right)}.
 	\end{equation}
Here, the different field contributions, including the saturation magnetization, are summarized in $M_\mathrm{eff, \parallel}= M_\mathrm{S}-H_\mathrm{u\perp}-H_\mathrm{c}$. Now the effective magnetization is obtained by fitting the $n=0$-mode (FMR mode). 

Secondly, the same equation is used to obtain the exchange field $\mu_0 H_\mathrm{ex}$ of the higher PSSW modes ($n \neq 0$). For this, the obtained value of $M_\mathrm{eff,\parallel}$ is used as a constant during the data evaluation process. The resulting exchange fields $\mu_0 H_\mathrm{ex}$ of the PSSW modes depend on the square of a mode number which is unknown at first. As in the case of the out-of-plane measurements there is an ambiguity regarding the identification of the mode number $n$ for each observed mode. The identification procedure of the mode numbers is shown next. 

Thirdly, the exchange fields of the modes are varied manually over the presumed mode numbers (see Fig.~\ref{klingler:fig:spectrum2}(b)). As first indicator, the peak height of the resonance can be used to determine whether a mode is an even mode or not. To proove the mode numbering a graphical feedback can be obtained by plotting $\mu_0 H_\mathrm{ex}(n)$ over $n^2$, where a wrong mode numbering would be directly visible. If the exchange field follows a linear function, as shown in Fig.~\ref{klingler:fig:spectrum2}(b), the exchange stiffness is given by the slope of this function.

For all three samples the values of the exchange stiffnesses are shown in the right column of Tab. \ref{klingler:tab:results}. They are in the same range with the other methods which supports the value of this method. However, the data evaluation is much more complicated and the systematic uncertainties are increased in comparison to the other methods. % It has to be emphasized that the statistical error of the in-plane measurements in Tab. \ref{klingler:tab:results} is the statistical error of the last step of the data evaluation process. Thus, it can be misleading in direct comparison of the results.

\section{Determination of the exchange constant}

\begin{table}
\caption{\label{klingler:tab:exchange_constants}Exchange constants of the YIG samples are shown for different methods. The proposed modified method of Schreiber and Frait (middle column) with excluded influence of anisotropies and $M_\mathrm{S}$ gives the best agreement for different samples.}
 \renewcommand{\arraystretch}{1.1}% Wider 
\begin{indented}
\item[]\begin{tabular}{@{}cccc}
\mr
		&  \multicolumn{3}{c}{\textit{A}\,$(10^{-12}\,$J/m) or (10$^{-7}$\,erg/cm)} \\
Sample	&\textit{out-of-plane$^\dagger$} & \textit{out-of-plane$^*$}& \textit{in-plane}\\
  		& \footnotesize{Schreiber and Frait}& &\\
\mr
 E1 & $3.64 \pm 0.40$	& $3.65 \pm 0.38$ & $3.71 \pm 0.39$ \\
 E2	& $3.64 \pm 0.43$	& $3.65 \pm 0.38$ & $3.63 \pm 0.38$ \\
 E3	& $3.76 \pm 0.44$	& $3.66 \pm 0.37$ & $3.73 \pm 0.40$ \\
\mr
\end{tabular}
\end{indented}
\end{table}

For the determination of the exchange constant $A=D M_\mathrm{S}/ 2$ of our YIG samples, the saturation magnetization $M_\mathrm{S}$ must be known. Vibrating sample magnetometry (VSM) was used to define $M_\mathrm{S}$ and values of 141\,kA/m, 136\,kA/m and 137\,kA/m for samples E1, E2 and E3, respectively, are found with an accuracy of $10\,\%$. The large error is due to the error in volume determination of the YIG films. The results obtained for $A$ using different methods of the definition of $D$ are shown in Tab.~\ref{klingler:tab:exchange_constants}. One can see that all values agree within the error bars. However, only the proposed out-of-plane$^*$ method gives practically the same value for all samples. This is due to the increased accuracy in the definition of the exchange stiffness. The exchange constant of the YIG films is determined to be $(3.65 \pm 0.38) \,$pJ/m, which is the average value of the out-of-plane$^*$ method. The presented result is in good agreement with the values obtained by other groups \cite{hoekstra}. Finally, one can state that the YIG films have the same material characteristics independent on the lattice mismatch and thickness, which speaks for the high quality of the YIG films (see Tab.~\ref{dubs:samples}).

\section{Conclusion}

Different methods were developed and compared to estimate the exchange stiffness $D$ from the microwave absorption spectra. Firstly, the method of Schreiber and Frait~\cite{frait} was used to estimate $D$ of the out-of-plane magnetized sample. The method was shown to be influenced by anisotropy fields and the saturation magnetization choice. Therefore, the exchange stiffnesses $D$ were accompanied with appreciable errors which resulted in different values for the exchange constant $A$. This problem was solved with the proposed method by avoiding additional fit parameters including anisotropy fields by preliminary extraction of the pure exchange field contributions [see Eq.~(\ref{eq:exchange_contribution})]. The method was demonstrated to give more accurate results which is the reason for the similar values of the exchange constant $A$. The modified method is recommended for determination of the exchange stiffness in general. Finally, the former method of was adapted for the in-plane configuration. The in-plane estimates of the exchange stiffness $D$ were found to agree well with those obtained in the out-of-plane configuration. However, because of the nonlinear dependence of the PSSW mode frequency versus $n^2$, an evaluation of the in-plane measurement data was more complicated and resulted in a similar spread of the exchange constant $A$ as in the original method of Schreiber and Frait \cite{frait}. 

Finally, it was also proven that the exchange constant in thin YIG films remain nearly independent of the YIG/GGG lattice misfit and a value of $A=(3.65 \pm 0.38)$\,pJ/m was extracted.

\section{Acknowledgements}
We thank the \textit{Nano Structuring Center} in Kaiserslautern for technical support. Part of this work was supported by NSF-CAREER Grant $\#0952929$ and by EU-FET grant InSpin 612759. The measurements were performed during the annual MINT Summer Internship Program of the University of Alabama.
\\


\begin{thebibliography}{}
\bibitem{DubsExplain} Pb contents in formular units were estimated from the lattice misfit by comparison with literature data \cite{Sure}.
\end{thebibliography}
\end{document}